\documentclass[a4paper]{aa}
\usepackage{txfonts,graphicx,natbib}
\bibpunct{(}{)}{;}{a}{}{,}
\newcommand{\pq}{\ensuremath{P_Q}}
\newcommand{\pu}{\ensuremath{P_U}}
\newcommand{\pv}{\ensuremath{P_V}}
\newcommand{\nv}{\ensuremath{N_V}}
\newcommand{\bz}{\ensuremath{\langle B_z\rangle}}

\newcommand{\nz}{\ensuremath{\langle N_z\rangle}}

\newcommand{\fo}{\ensuremath{f^\parallel}}
\newcommand{\fe}{\ensuremath{f^\perp}}
\begin{document}
\title{The importance of non-photon noise\\ in stellar spectropolarimetry}
\subtitle{\it The spurious detection of a non-existing magnetic field in the A0 supergiant HD\,92207}
       \author{
        S.~Bagnulo      \inst{1}
       \and
        L.~Fossati      \inst{2}
       \and
        O. Kochukhov    \inst{3}
       \and
        J.D.~Landstreet  \inst{1,4}
        }
\institute{
           Armagh Observatory,
           College Hill,
           Armagh BT61 9DG,
           Northern Ireland, U.K.
           \email{sba@arm.ac.uk}
           \and
           Argelander-Institut fuer Astronomie der Universitaet Bonn,
           Auf dem Huegel 71, D-53121, Bonn, Germany.\\
           \email{lfossati@astro.uni-bonn.de}
           \and
	   Department of Physics and Astronomy, 
	   Uppsala University, SE-751 20,
	   Uppsala, Sweden.\\
	   \email{oleg.kochukhov@physics.uu.se}
           \and
           Physics \& Astronomy Department,
           The University of Western Ontario,
           London, Ontario, Canada N6A 3K7.\\
           \email{jlandstr@uwo.ca}
}

\authorrunning{S.\ Bagnulo et al.}
\date{Received: 2013-07-18 / Accepted: 2013-08-26}
\abstract
{
The low-resolution, Cassegrain mounted, FORS spectropolarimeter of the
ESO Very Large Telescope is being extensively used for magnetic field
surveys. Some of the new discoveries suggest that relatively strong
magnetic fields may play an important role in numerous physical
phenomena observed in the atmospheres as well as in the circumstellar
environments of certain kinds of stars.
}
{
We show in detail how small instabilities or data-reduction
inaccuracies represent an alternative explanation for the origin of
certain signals of circular polarisation published in recent years.
}
{ 
With the help of analytical calculations we simulate the observation
of a spectral line in spectropolarimetric mode, adding very small
spurious wavelength shifts, which may mimic the effects of seeing
variations, rapid variations of the stellar radial velocity, or
instrument instabilities. As a case
study, we then re-visit the FORS2 measurements that have been used
to claim the discovery of a magnetic field in the A0 supergiant
HD\,92207. In addition, we present new observations of this star
obtained with the HARPSpol instrument.
}
{
Both calibration and science data show compelling evidence that
photon-noise is not the only source of error in magnetic field
measurements, especially in sharp spectral lines. Non-photon
noise may be kept under control by accurate data reduction and
quality controls.  Our re-analysis of FORS2 observations of HD\,92207
shows no evidence of a magnetic field, and we are able to
reproduce the previous FORS detection only by degrading the quality of our
wavelength calibration. Our HARPSpol spectropolarimetric measurements
show no evidence of a magnetic field at the level of 10\,G.
}
{
Our work contributes to a better understanding of the importance of
accurate data treatment and instrument characterisation, and demonstrates that ultra-high
signal-to-noise ratio measurements do not automatically translate into
ultra-high accuracy.
}
\keywords{polarisation -- 
techniques: polarimetric --
stars: magnetic field -- Stars: individual: HD 92207}
\maketitle
\section{Introduction}
Polarimetry is attracting a steadily growing level of interest in
various research areas of astronomy. The most popular and successful
application of spectropolarimetery is to magnetic stars, but
polarimetric techniques may make significant contributions to the
advance of knowledge in many other fields. Perhaps not surprisingly,
the most interesting applications are those that require measurements
at the very limit of the present and near future instrumentation, for
example the characterisation of the atmospheres of exo-solar planets,
and the search for extra-terrestrial life \citep{Steretal12}.

Even in the traditional area of stellar magnetism, much of the most
interesting work recently has been directed to the detection of very
weak fields in a variety of stars that in the past were thought to be
non-magnetic, or in which fields were undetectable. Fields have
recently been discovered in a variety of types of stars such as Of?p
stars \citep[][and references therein]{GruWad12}, $\beta$~Cep and
Slowly Pulsating B stars \citep{Henetal00,Neietal03,Siletal09}, Herbig
AeBe stars \citep{Wadetal05,Wadetal07,Catetal07,Aletal13}, red giant
\citep{Auretal09} and AGB supergiant stars \citep{Gruetal10}, and many
kinds of cool dwarfs \citep[][and references therein]{Donetal09}.
Fields in such stars have become detectable as a result of having new
generation instrumentation of very high efficiency and broad
wavelength coverage available on medium
and large telescopes, and through advanced data reduction and treatment
techniques that allow very small polarimetric signatures to be
reliably detected. However, at the same time that these new techniques
have made possible the discovery of magnetic fields in many stars, the
difficulties of reducing the data and evaluating the uncertainties of
measurement correctly have led to a number of reports of discoveries
of fields in stars in which fields have not in fact been detected
\citep{Bagetal12}.

The aim of this paper is to discuss how easily spurious signals can
appear in polarimetric observations, and to discuss some general
considerations that may be relevant when designing new instrumentation
and reduction pipelines for high-precision spectropolarimetry. We will
try to clarify some specific sources of error and uncertainty in
high-precision spectropolarimetry that we believe have contributed
significantly to previous incorrect reports of field detections. This
work may therefore be applicable to a wider domain than simply that of
stellar magnetism, because of the expanding relevance of
spectropolarimetry (including linear spectropolarimetery) to many
domains. We hope that this discussion may contribute to significantly
reducing the number of dubious or erroneous reports of field
discoveries appearing in the literature.
\begin{figure*}
\scalebox{0.26}{
\includegraphics*[1.2cm,5cm][25.5cm,26cm]{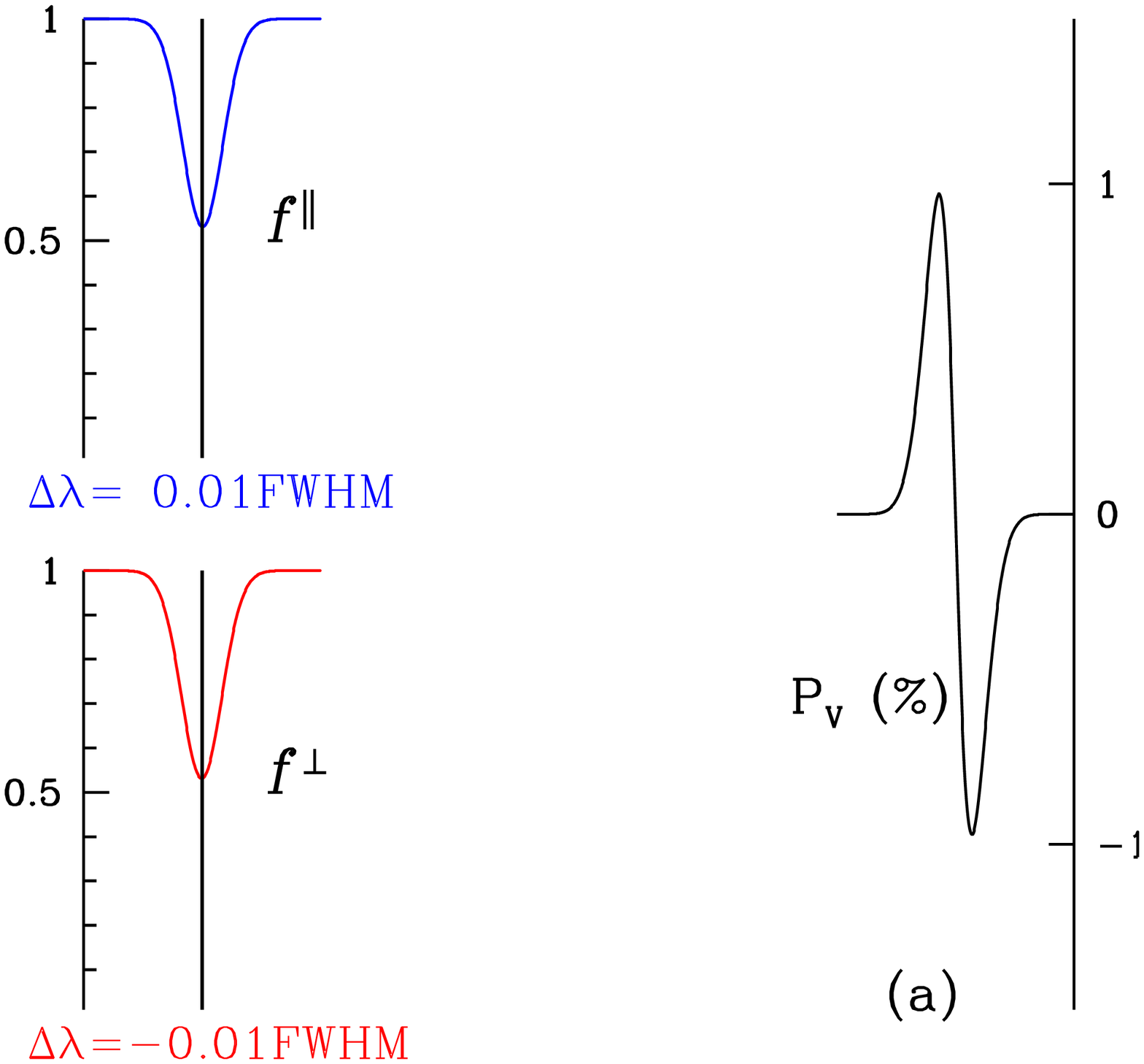}}
\scalebox{0.26}{
\includegraphics*[0cm,5cm][25.5cm,26cm]{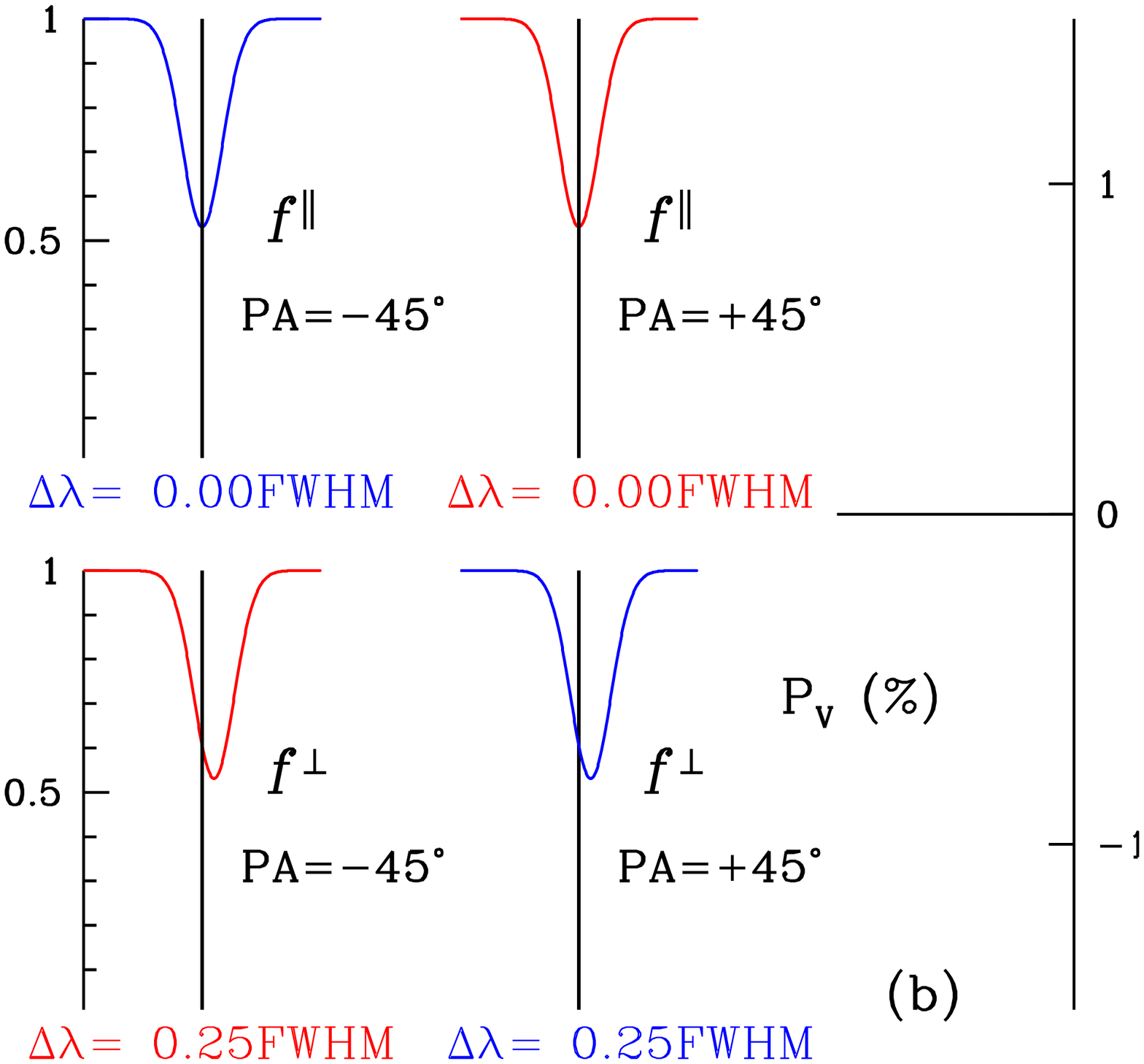}}
\scalebox{0.26}{
\includegraphics*[0cm,5cm][25.5cm,26cm]{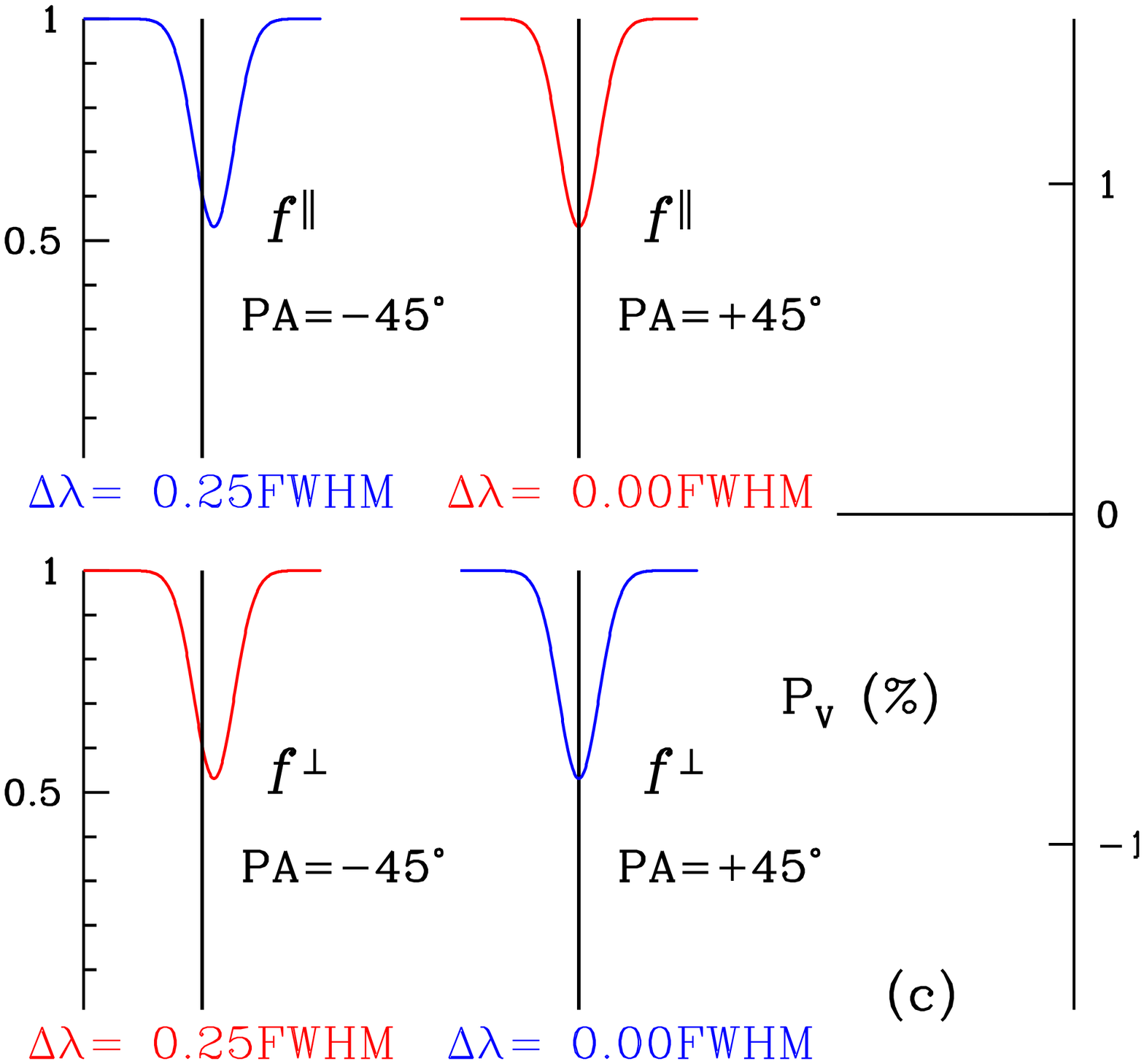}}\\
\scalebox{0.26}{
\includegraphics*[1.2cm,5cm][25.5cm,26cm]{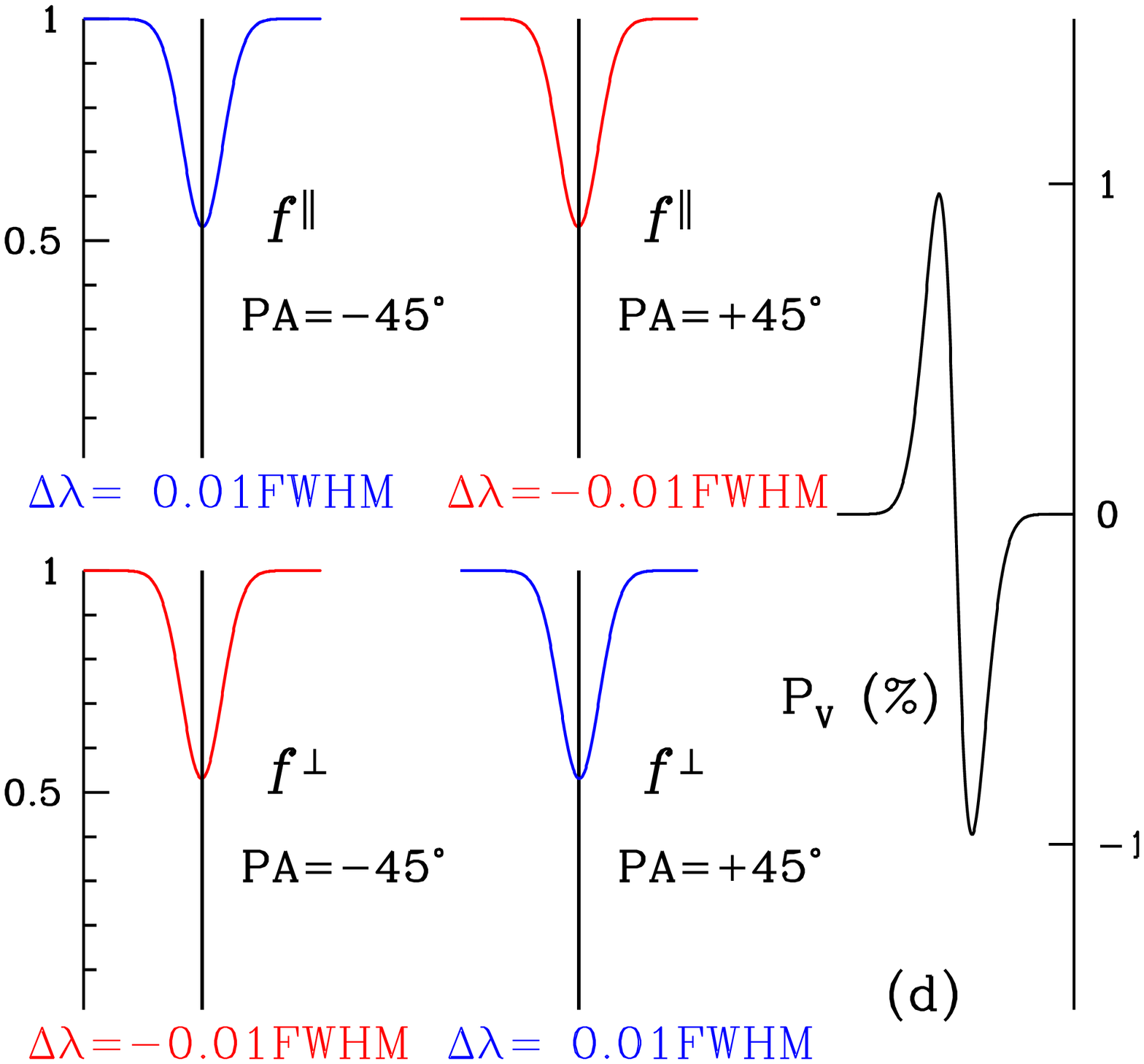}}
\scalebox{0.26}{
\includegraphics*[0cm,5cm][25.5cm,26cm]{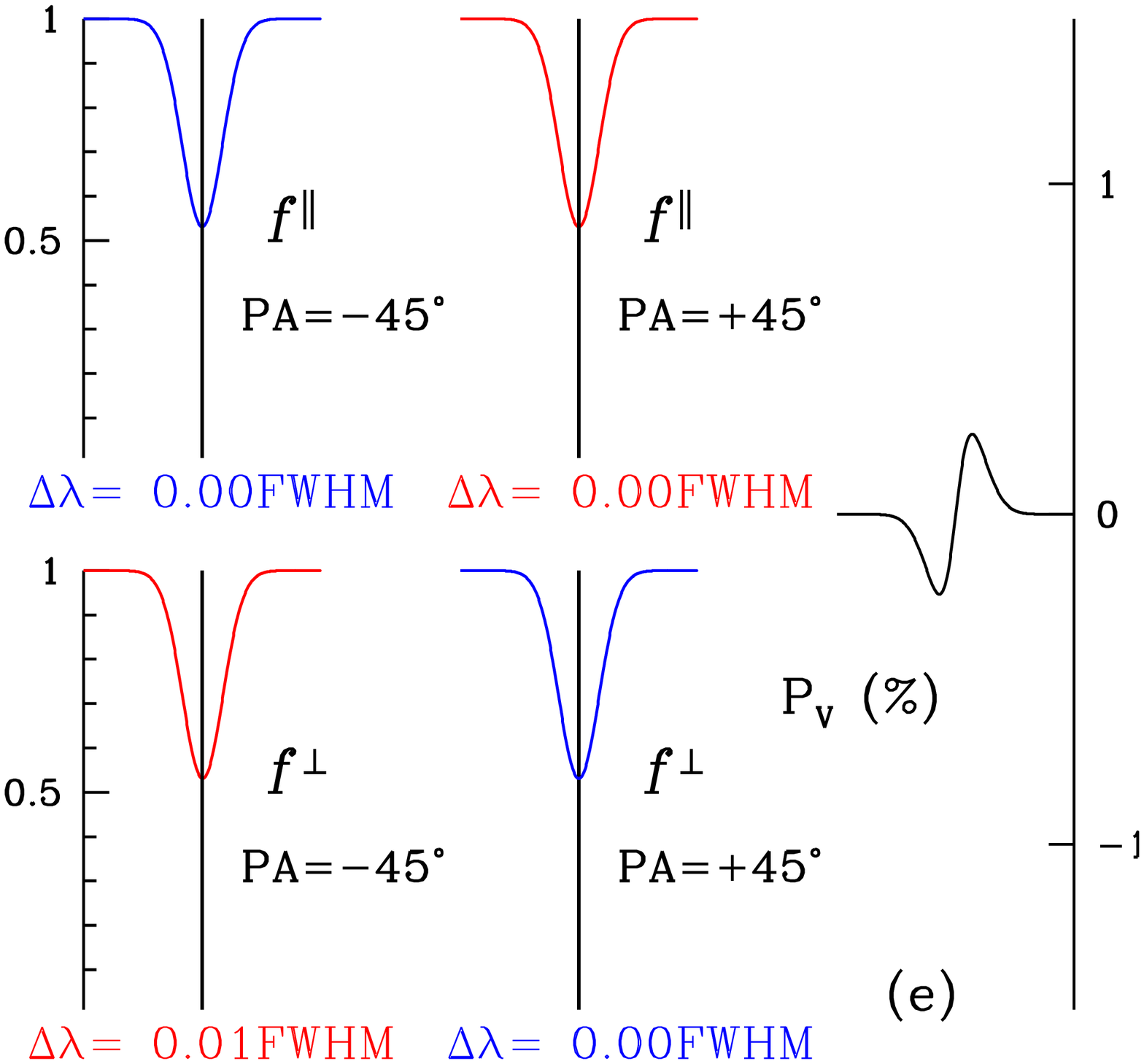}}
\scalebox{0.26}{
\includegraphics*[0cm,5cm][25.5cm,26cm]{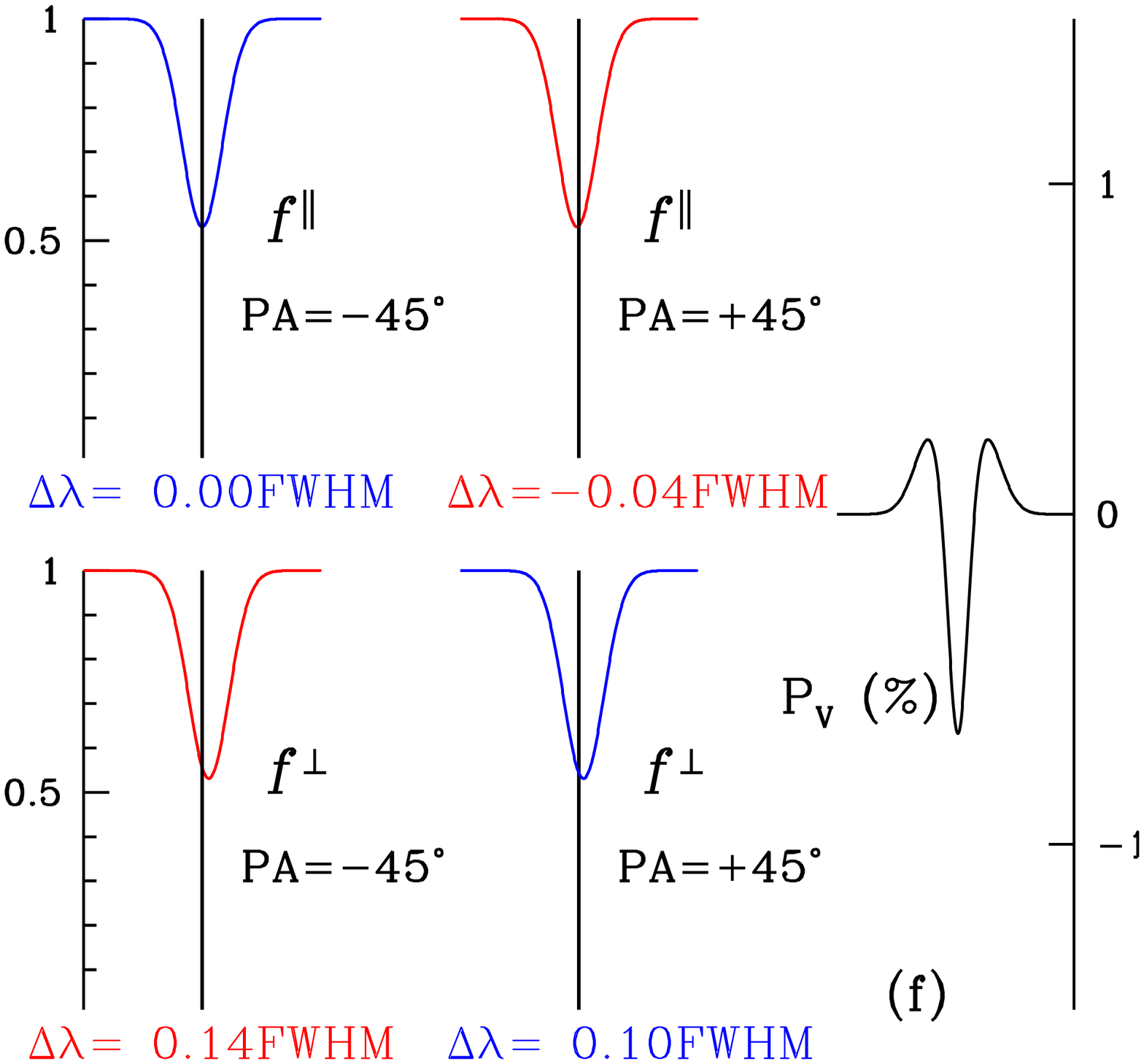}}
\caption{\label{Fig_Examples} Combinations of Gaussian profiles to
simulate polarimetric observations of a spectral line. The Figure is explained in the text.}
\end{figure*}

In this paper we present in Sect.~\ref{Sect_Simulations} a set of simple numerical
simulations that illustrate some important issues about instrument
stability and wavelength calibration. In Sect.~\ref{Sect_Offsets} we analyse some
calibration and science data obtained with the FORS low-resolution
spectropolarimeter of the ESO VLT, which is one of the most active
instruments in the search for stellar magnetism. These data allow us
to estimate the size of some kinds of data uncertainty in observations.
In Sect.~\ref{Sect_HD92207} we re-discuss FORS
observations of the A0 supergiant star HD\,92207 that have led to the
claim by \citet{Hubetal12} of the discovery of a new class of magnetic
stars, and we present new observations of HD\,92207 obtained with
the HARPSpol spectropolarimeter at the 3.6\,m telescope of the ESO La
Silla Observatory. In Sect.~\ref{Sect_Discussion} we discuss our results
and present our conclusions.

\section{The effects of the beam-swapping technique on wavelength shifts}\label{Sect_Simulations}
There are different physical mechanisms that may be responsible for
the circular polarisation of a spectral line. The Zeeman effect causes
a differential offset of a spectral line observed in opposite
polarisations.

To simulate the effects of small instrument flexures during the
observation of a spectral line in polarimetric mode, we may combine
Gaussian profiles after they are offset by a small fraction of their
FWHM. Figure~\ref{Fig_Examples} shows various examples of such
combinations, considering Gaussians with FWHM=1 and depth = 0.5.  Case
(a) is the simple combination of two Gaussian profiles offset by $\pm
0.01$\,FWHM; the profile on the right-hand side of panel (a) is given
by the ratio between the difference and the sum of the two profiles on
the left-hand side, $(\fo-\fe)/(\fo+\fe)$, expressed in percent
units. This case may represent a polarised spectral line observed in
the parallel (\fo, upper line) and in the perpendicular (\fe, lower
line) beam after the light has travelled across a retarder waveplate
and a Wollaston prism.  In case of circular polarisation, the parallel
beam has the right-hand circularly (RHC) polarised profile, the
perpendicular beam has the left-hand circularly (LHC) polarised
profile (in all panels of Fig.~1, the RHC flux is represented by blue
lines, and the LHC flux is represented by red lines).  This case could
also represent the observation of a non polarised spectral line where
the offsets have been artificially introduced by e.g., a slightly
incorrect wavelength calibration of the parallel and perpendicular
beams.

In the remaining cases we simulate the \textit{beam swapping}
technique, which is commonly employed in real polarimetric
observations. If the $\lambda/4$ retarder waveplate is rotated by
90\degr, then the signal in the parallel beam becomes proportional to
the LHC polarised light, and the signal in the perpendicular beam is
proportional to the RHC polarised light. The reduced $\pv = V/I$
profiles are obtained with the formula
\begin{equation}
\pv = \frac{1}{2} \left[ 
\left(\frac{\fo - \fe}{\fo + \fe}\right)_{-45^\circ} - 
\left(\frac{\fo - \fe}{\fo + \fe}\right)_{+45^\circ}\right]\,.
\label{Eq_V}
\end{equation}
The advantages of this technique are that many spurious effects cancel
out, as originally explained in the FORS user manual, and explained in
greater detail by \citet{Bagetal09}.  Case (b) of
Fig.~\ref{Fig_Examples} shows that if the wavelength shift between
parallel and perpendicular beam is an instrumental or a calibration
artifact which {\it remains constant as the retarder waveplate
  rotates}, then its effect is canceled out if the fluxes are combined
as prescribed by Eq.~(\ref{Eq_V}), and the resulting reduced Stokes
profile is flat. Similarly, the effect of an offset that is present in
both parallel and perpendicular beams only at one position of the
retarder waveplate (case c) is canceled out.  If the wavelength shift
is due to the polarisation intrinsic to the spectral line, then the
offset between top and bottom beam is reversed when the retarder
waveplate is rotated by 90\degr\ (case d), and the combination given
by Eq.~(\ref{Eq_V}) reproduces a genuine Zeeman feature. Finally,
cases (e) and (f) show that an artificial offset between spectral
lines in opposite polarisations, which appears or changes after the
retarder waveplate rotates, may be responsible for a spurious
polarisation signal.

\section{The occurrence of spurious offsets and their impact on science data}\label{Sect_Offsets}
To discuss the likelihood of the occurrence of spurious offsets in real
observations, we need to investigate the spectral image stability and
compare it to the FWHM of the spectral lines used for magnetic field
detection.

\subsection{Offsets introduced by the polarimetric optics}\label{Sect_Pol_Off}
To have an approximate idea of the offsets potentially involved in
practical applications, we first consider calibration data
obtained with the FORS instrument in spectropolarimetric mode, and
we estimate the size of the offsets introduced by the
  polarimetric optics. In particular, we inspect the images of the
arc lines used for wavelength calibration with the grating 600\,B,
when the CCD is read out with no rebinning. Pixel size is
15\,$\mu$m, which corresponds to a pixel scale of 0.125\arcsec\ on sky.  Instrument
dispersion is $\sim 0.75$\,\AA\ per pixel. All data considered in the
paper were obtained with the E2V CCD mosaic, but the geometrical
characteristics are identical for the more commonly used MIT CCD
mosaic.

We recall that calibrations are taken during day time with the
telescope at zenith, using the inner side of the Cassegrain shutter as
a screen. This implies that slit illumination is constant from one
calibration arc to the next, which is not the case for stellar
observations.

\begin{figure}
\scalebox{0.60}{
\includegraphics*[3.5cm,9cm][17.7cm,20.5cm]{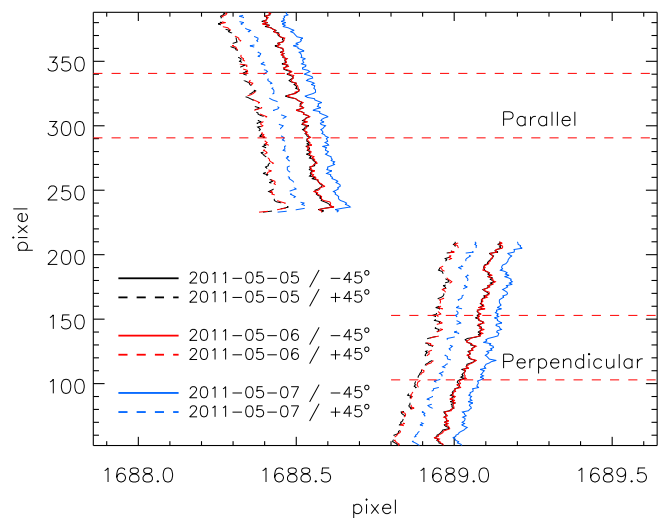}}

\caption{\label{Fig_Tracing} Tracing of the centre of the \ion{Hg}{i}
  arc line at $\lambda=4358.343$\,\AA\ on the FORS2 detector in the
  spatial region where science spectra are recorded. Different colours
  and line-styles refer to calibration frames obtained on different
  days and with different positions of the retarder waveplate, as
  explained in the text. Note that black and red lines are practically
  superposed, but the line positions on 2011-05-07 are offset by $\sim 0.08$\,px
with respect to those measured on 2011-05-05 and 2011-05-06.}

\end{figure}

We look at the 400 lower rows of the upper chip, which is
the region used for most of the science observations when only one
target is observed. Specifically, the perpendicular beam is recorded
in the region around pixel lines $\sim 100-150$ pixels, while the
parallel beam is recorded in the region around pixel lines $\sim
290-340$. Figure~\ref{Fig_Tracing} shows a detail of the tracing of
the arc \ion{Hg}{i} line at $\lambda=4358.343$\,\AA, obtained in
different exposures with the retarder waveplate at position angles
$-45\degr$ and $+45\degr$, and on three consecutive days. The tracing was
carried out by fitting the line profile in each pixel row (the
$x$-axis of the CCD) with a Gaussian profile, and then by following
the Gaussian peaks along the direction perpendicular to the dispersion
(the $y$-axis of the CCD).  The arc line FWHM is 3.3 pixels, for a
spectral resolution of $\Delta \lambda = 2.5$\,\AA.

It appears that, due to field distortion, and in particular due to the
fact that the beams split by the Wollaston follow different optical
paths, the perpendicular (bottom) beam is redshifted by approximately
0.55\,px (i.e., about 15\,\% of the line FWHM) with respect to the
parallel (top) beam. When the retarder waveplate is rotated from
$-45\degr$ to $+45\degr$ all beams are blue-shifted by $\sim 0.13$\,px
($\sim 4$\,\% of the line FWHM). This very small offset is likely due
to the fact that the retarder waveplate does not rotate in a plane
perfectly perpendicular to the incoming light beam. Line positions are
approximately, but not perfectly, reproduced when the retarder
waveplate is rotated, removed and re-inserted at different epochs --
this is likely due to a combination of flexures/hysteresis of the
system telescope+instrument.
Such offsets, if not corrected, would generate a spurious signal of
polarisation qualitatively similar to that of Fig.~\ref{Fig_Examples}
(case f). 

\begin{figure}
\begin{center}
\scalebox{0.56}{
\includegraphics*[2.5cm,8.6cm][19.8cm,21.5cm]{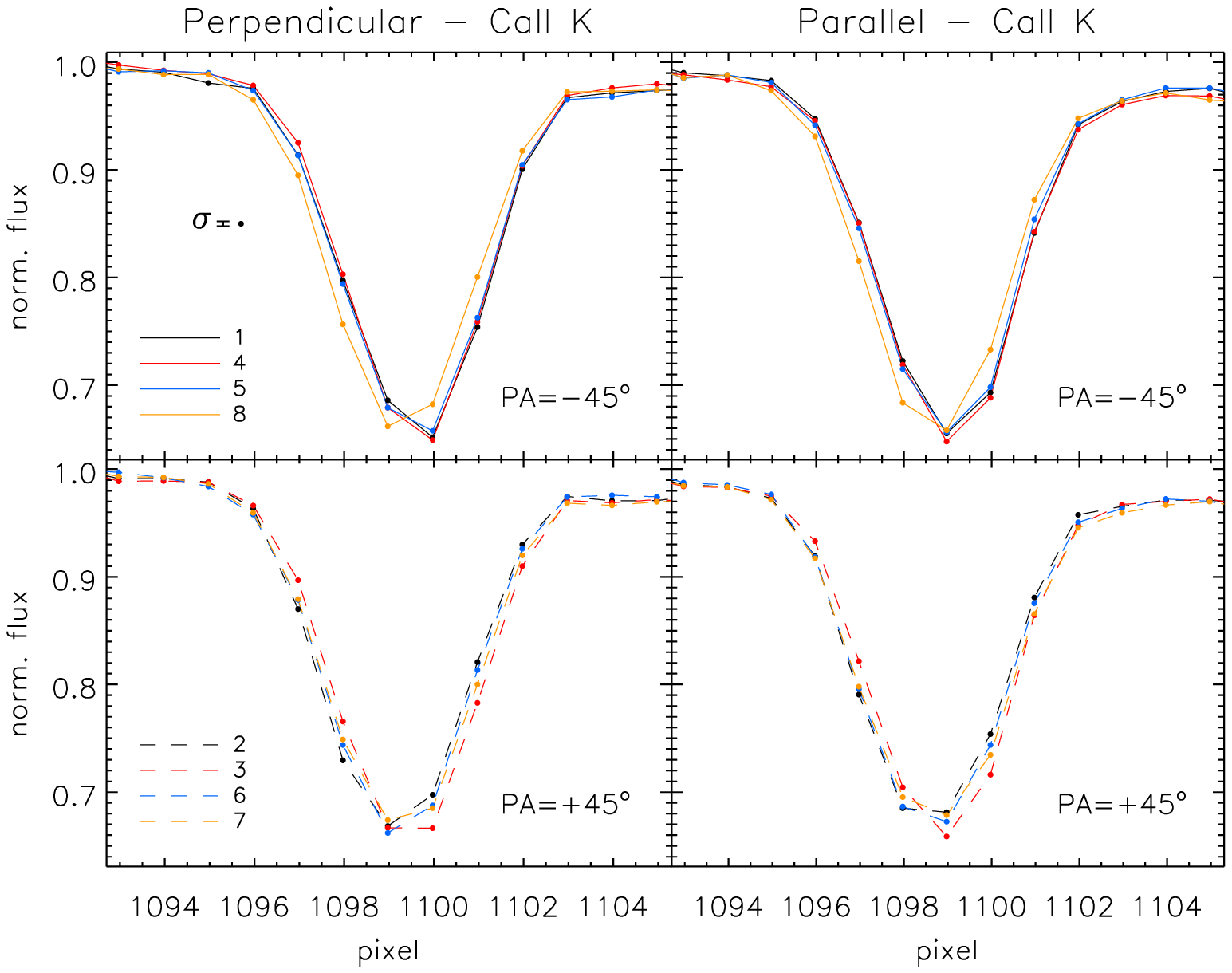}}\\
\scalebox{0.56}{
\includegraphics*[2.5cm,8.6cm][20cm,21.5cm]{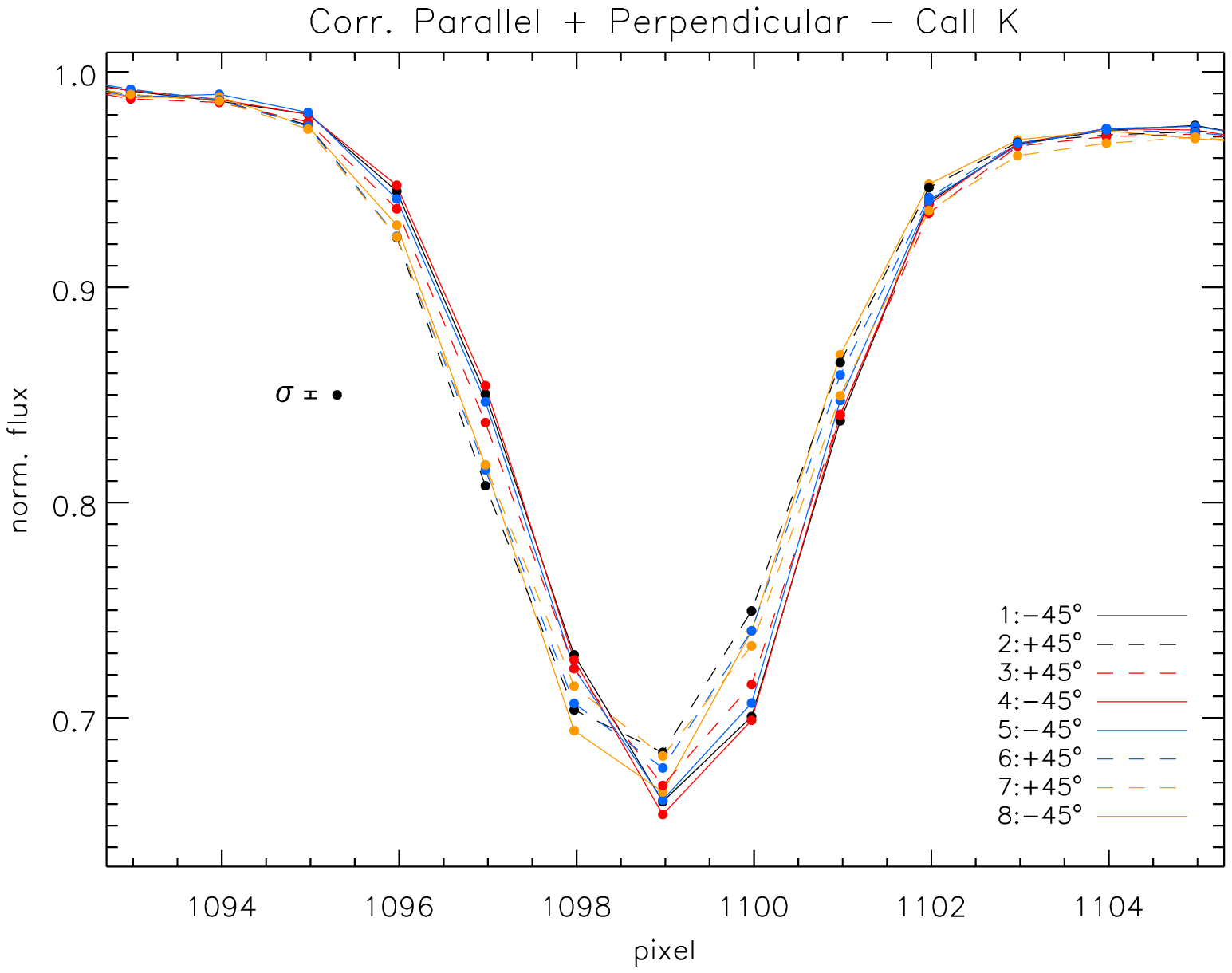}}
\end{center}
\caption{\label{Fig_CaII_2011} Profiles of the \ion{Ca}{ii}\,K line extracted
from the FORS frames of the night on 2011-05-07 as follows:
1: UT 03:56:53 (-45\degr); 
2: UT 03:58:28 (+45\degr); 
3: UT 03:59:32 (+45\degr); 
4: UT 04:01:08 (-45\degr); 
5: UT 04:02:12 (-45\degr); 
6: UT 04:03:49 (+45\degr); 
7: UT 04:04:54 (+45\degr); 
8: UT 04:06:29 (-45\degr).
Note that pairs (2,3), (4,5) and (6,7) were obtained consecutively without
changing the instrumental setup. The retarder waveplate is rotated after the end of the
exposures No.\,1, 3, 5, and 7.
The Figure is explained in the text.
}
\end{figure}
\begin{figure}
\begin{center}
\scalebox{0.54}{
\includegraphics*[2.5cm,8.6cm][20cm,21.5cm]{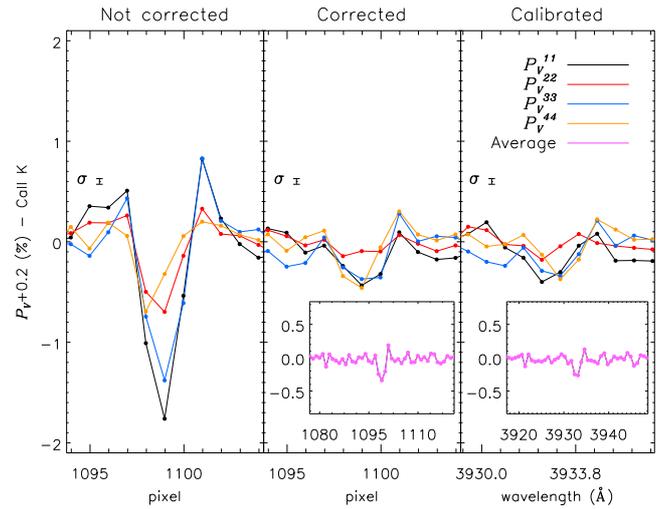}}
\end{center}
\caption{\label{Fig_CaII_Cal_2011} \pv\ profiles of the \ion{Ca}{ii}\,K line of
the night 2011-05-07 combined as explained in the text.
$\pv^{11}$: UT 03:56:53 (-45\degr) + UT 03:58:28 (+45\degr); 
$\pv^{22}$: UT 04:01:08 (-45\degr) + UT 03:59:32 (+45\degr); 
$\pv^{33}$: UT 04:02:12 (-45\degr) + UT 04:03:49 (+45\degr);  
$\pv^{44}$: UT 04:06:29 (-45\degr) + UT 04:04:54 (+45\degr).
The inset provide an enlarged view of the average of the 
$\pv^{jj}$ profiles of the \ion{Ca}{ii} line.
}
\end{figure}
The risk of spurious artefacts may be minimised with a very accurate
wavelength calibration, and certainly, accurate wavelength calibration
may compensate for the offset between parallel and perpendicular beam.
To take into account the shifts introduced when the retarder waveplate
rotates, one could calibrate the science observations obtained with
the retarder waveplate at a certain position angle using the arc frame
obtained with the retarder waveplate at the same position angle,
making sure that arc frames at $-45\degr$ and $+45\degr$ are obtained
without having the retarder waveplate removed and re-inserted in
between. This means that each observing set needs to be treated with
at least four different wavelength calibration solutions. However,
experience shows that the additional noise introduced by adopting two
independent calibration frames may be responsible for larger
artificial offsets than those that they are meant to be corrected
\citep{Bagetal06,Bagetal09}, potentially leading to spurious
detections, such as those discussed by \citet{Joretal12} and
\citet{Lanetal12}. Of course, two independent solutions must be
obtained for the two beams split by the Wollaston prism, and to keep
the highest level of consistency, these two solutions must be obtained
adopting exactly the same fitting function and using exactly the same
arc lines.  Once the offset between parallel and perpendicular beam is
corrected with two independent wavelength calibrations, then we are in
case (c) of Fig.~\ref{Fig_Examples}, and the effect of the offset
introduced when the retarder waveplate rotates should be fully
compensated. Alternatively, one could simply apply a rigid offset of
$0.13$\,px to all frames obtained with the retarder waveplate at
$+45\degr$ (which corresponds to a wavelength step that depends on the
specific instrument setting). However, since day-time calibrations do
not reproduce the exact optical path of the night-time science
observations, we cannot confirm that is the best data reduction
strategy.

\subsection{Offsets appearing in science spectra}
As a real science application, in this Section, and in
Sects.~\ref{Sect_FORS_data} and \ref{Sect_Null}, we examine in detail
the magnetic observations of the A0 supergiant HD\,92207, for which
\citet{Hubetal12} have claimed discovery of a magnetic field on the
basis of two apparently significant field detections, one of $-402 \pm
52$\,G and one of $+157 \pm 51$\,G, using Balmer lines (slightly
different results are obtained from the analysis of the full
spectrum). We focus on the first of these two measurements, which is
nominally significant at the $8\,\sigma$ level. These observations
were obtained on 2011-05-07 with the FORS2 instrument of the ESO
VLT. The grism 600\,B was used with an 0.4\arcsec slit width.  FORS2
data reduction was performed independently by two of us (SB and LF)
using various IRAF tasks and independent FORTRAN and IDL routines, and
repeated using the ESO pipeline \citep{Izzetal10}.

Various simple tests can be devised to check the stability of the spectrum image. For
instance, in the ideal case, a given line profile (normalised to the
continuum) of a non variable star, obtained under identical instrument
settings, should appear constant within photon-noise error bars. The
visual impact of a wavelength shift is clearly higher on deep and
sharp lines than on shallow lines. Here we consider the
\ion{Ca}{ii}\,K line at $\lambda = 3933.7$\,\AA, which, at the
instrument resolution, appears as a deep and sharp feature, with
a FWHM of about 3.6 pixels. 

The four top panels of Fig.~\ref{Fig_CaII_2011} show the profiles
extracted from the parallel and perpendicular beams obtained at two
positions of the retarder waveplate, prior to wavelength calibration.
Different colours and line-styles are used for the profiles extracted
during different exposures obtained in temporal sequence as given in
the caption. Ideally, all profiles within each panel should coincide
within photon-noise error bars, which is about the same size as the
adopted symbols. However, even profiles obtained consecutively without
moving the retarder waveplate, e.g. profiles 2 and 3, appear clearly
shifted. The largest offset is the one observed between profiles 5
and 8, both obtained with the retarder waveplate at $-45\degr$ (two upper
panels). The centre of the Gaussian used to approximate profile no.\,8
is blue-shifted by 0.25\,px with respect to the Gaussian that best fits
the profile no.\,5.

The bottom panel of Fig.~\ref{Fig_CaII_2011} shows the corresponding
Stokes $I$ profiles. The profiles extracted in the parallel and in the
perpendicular beam have been added together, after correcting for the
shifts observed in Fig.~\ref{Fig_Tracing}, namely: all perpendicular
beams have been offset by $-0.55$ pixels and all beams obtained with
the retarder waveplate at $+45\degr$ have been offset by $+0.13$\,px.
In the ideal case, regardless the presence of a magnetic field, all
these profiles should be indistinguishable within photon-noise error
bars. In contrast, the obvious presence of shifts shows that either
the instrument or the star image on the slit are stable only within a
fraction of a pixel (say 25\,\%, corresponding to $\sim
0.03$\arcsec\ on sky). Our conclusions are that spectral lines in
  consecutive science exposures have an offset pattern more complex
  than that observed in the arc lines as the retarder waveplate
  rotates, and that the actual error bars on spectral lines are
larger than those due to Poisson noise. This is consistent with and
further supports a general finding by \citet{Bagetal12}.

It is of some interest to display the combinations of the profiles of
Fig.~\ref{Fig_CaII_2011} according to Eq.~(\ref{Eq_V}). The leftmost
panel of Fig.~\ref{Fig_CaII_Cal_2011} shows the \ion{Ca}{ii}\,K
profiles combined without wavelength calibration; the result is a deep
spike with a $\sim 2$\,\% amplitude, qualitatively similar to panel
(f) of Fig.~\ref{Fig_Examples}. The middle panel shows the same
combination after applying the offest correction measured on the arc
line of Fig.~\ref{Fig_Tracing}. The feature in the ``pseudo-\pv''
profiles is now greatly reduced. Finally, the rightmost panel shows
the ``true'' \pv\ profiles, i.e., obtained after a global wavelength
calibration of the various profiles. A feature is still visible, but
from our previous analysis we conclude that it is unlikely that it is
due to the Zeeman effect. We finally note that a similar analysis
carried out for other spectral lines lead to similar conclusions as
that carried out for the \ion{Ca}{ii} line.

\section{The non detection of a magnetic field in the A0 supergiant HD\,92207}\label{Sect_HD92207}
\begin{figure}
\scalebox{0.45}{
\includegraphics*[0.5cm,5.5cm][22cm,25cm]{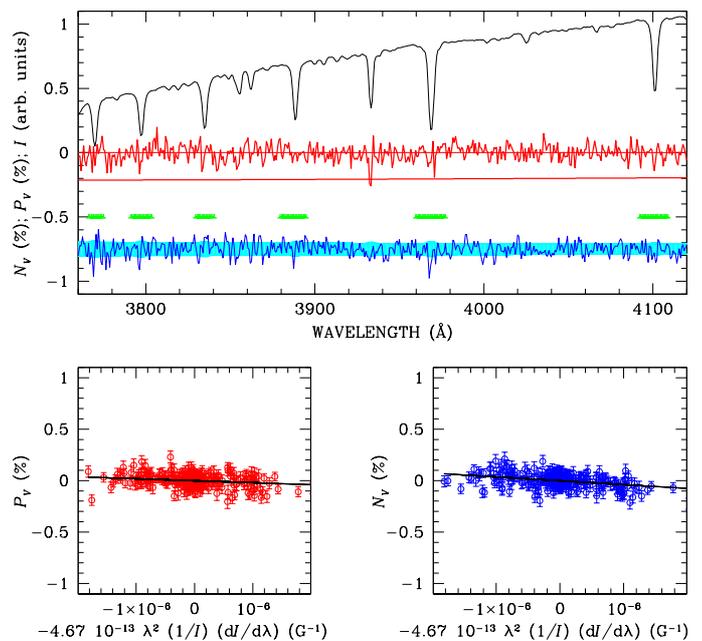}}
\caption{\label{Fig_HD92207_nospikes} Polarised spectrum of HD\,92207
observed on 2011-05-07 with grism 600B. The top panel shows 
the observed Stokes $I$ profile (black solid line, in arbitrary units,
and not corrected for the instrument response), the rectified \pv\
profile (red line centred about 0; the red smooth solid line slightly offset 
from zero shows the position of the \pv\ ``continuum'' prior rectification)
and the null profile (blue solid line, offset by $-0.75$\,\% for display
purpose). The \pv\ error bars are represented with a light blue band centred
about –0.75\,\%.
The slope of the interpolating lines in the bottom panels
gives the mean longitudinal field from \pv\ (left bottom panel) and
from the null profile (right bottom panel), both calculated using 
the H Balmer lines only. The regions adopted to derive the magnetic field are shown by 
thick green lines centered around $-0.5$. The corresponding \bz\ and \nz\ values are
$-190\pm65$\,G and $-375\pm65$\,G, respectively.}
\end{figure}
\begin{figure}
\scalebox{0.45}{
\includegraphics*[0.5cm,5.5cm][22cm,25cm]{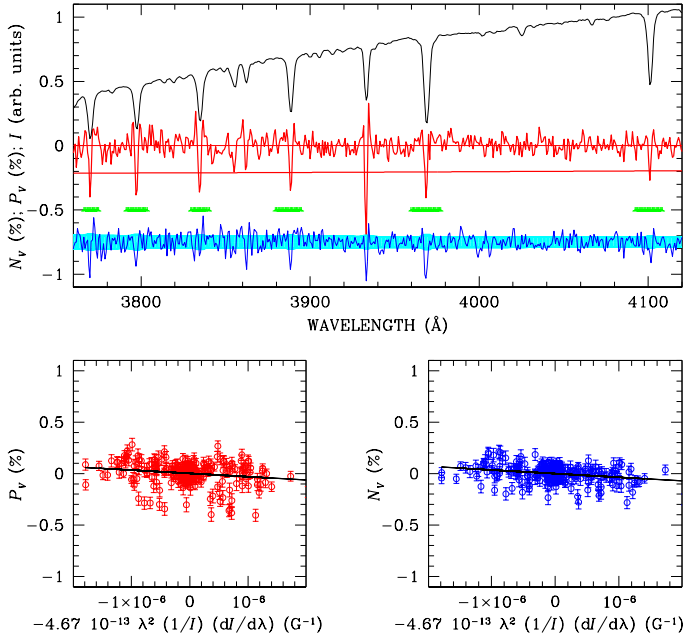}}
\caption{\label{Fig_HD92207_spikesVN}
As Fig.~\ref{Fig_HD92207_nospikes}, but adopting a less than optimal
wavelength calibration as explained in the text. Note that spikes
are detected both in \pv\ and in the \nv\ profiles. The corresponding
\bz\ and \nz\ values are $-325\pm105$\,G and $-355\pm75$\,G, respectively.
}
\end{figure}
\subsection{FORS archive data}\label{Sect_FORS_data}
The results of our reduction of FORS2 data (see
Fig.~\ref{Fig_HD92207_nospikes}) are significantly different from
those published by \citet{Hubetal12}. First of all, our circular
polarised profiles show a global offset of $\sim -0.2\,\%$. This could
be a symptom of cross-talk from linear to circular polarisation
\citep{Bagetal09}: a fraction of the linear polarisation intrinsic to
the source may be transformed into circular polarisation by the
telescope or instrument optics, before reaching the polarimetric
optics.  Based on FORS1 linear polarisation observations,
\citet{Ignetal09} reported for HD\,92207 an average continuum
polarisation of $\pq \sim -2$\,\% and $\pu \sim -2.6$\,\%. Therefore
the continuum of circular polarisation could be explained as a 10\,\%
cross-talk from \pq\ to \pv, which, although a bit higher than
expected, is not unrealistic. Another possible explanation (supported
by simple numerical simulations) is the presence of scattered light
that slightly changes from one exposure to the next.

Compared to the results by \citet{Hubetal12}, the most noticeable
difference is however that, once the profile is rectified, our field
measurement is consistent with zero: $\bz=-190\pm 65$\,G;
the rectification procedure, and the way the longitudinal field is
calculated, are described in detail by \citet{Bagetal12}, and the
error bar is given using their Eq.~(11), i.e., we have
considered the external error; however, in this paper, we have
\textit{not} performed any clipping on the data. Following some
tendency in the literature, one might be tempted to describe our
result as a $2.9\,\sigma$ probable detection. However, our analysis of
the profile instability presented in the previous Section strongly
suggests that this is instead simply a null detection, and that
also the feature associated to the \ion{Ca}{ii} \citep[visible in our
  plot, although less pronounced than that shown by][]{Hubetal12}, is
almost certainly spurious. A further indication that our ``2.9\,$\sigma$
detection'' should not be treated as a marginal detection
comes from the $5.7\,\sigma$ field detection in the null
profile: $\nz = -375\pm65$\,G (we recall that null profiles are expected
to be oscillating about zero within error bar, and that, if used to
measure the magnetic field, one should obtain a value consistent with
zero).  

It remains to understand why \citet{Hubetal12}, from the same dataset,
obtained an $~8\,\sigma$ detection. This measurement, if correct,
would represent a firm field detection (from very general statistical
considerations, \citet{Bagetal12} concluded that FORS magnetic field
detections are secure only if they reach at least a $5-6\,\sigma$
level). Aside from the fact that our data-reduction does not lead to a
field detection, we point out that the shape of the \pv\ profiles
shown in Fig.~1 of \citet{Hubetal12} can hardly be explained in terms
of Zeeman effect. Unless the region of line formation is characterised
by strong radial convective modes and the magnetic field changes with
optical depth, the \pv\ profiles formed in a magnetic atmosphere
always have the zero-order moment equal to zero (i.e., the integral of
Stokes $V$ over the region of one spectral line is equal to zero). In
contrast, the \pv\ profiles shown in Fig.~1 of \citet{Hubetal12} seem
to have a strongly negative zero-order moment, and qualitatively
resemble the \pv\ profiles obtained by combining the non
wavelength-calibrated beams shown in the left panel of our
Fig.~\ref{Fig_CaII_Cal_2011}. Therefore we suggest that the field
detection of \citet{Hubetal12} is the result of the combined effect of
a less than optimal wavelength calibration and a not perfectly stable
spectral image on the CCD.

To support our hypothesis, in Fig.~\ref{Fig_HD92207_spikesVN} we show
the results of a different reduction of the same data, in which we
have introduced a small difference in the way we have wavelength
calibrated the parallel and the perpendicular beam. For the wavelength
calibration of the parallel beam we have interpolated with a third order
cubic spline 22 lines of the arc spectrum listed in the line
catalogue of the FORS user manual. For the wavelength calibration of
the perpendicular beam we have used a first order cubic spline, and we
have not included two blue \ion{He}{i} arc lines, one at
$\lambda=3888.6$\,\AA\ and one at $\lambda=4026.1$\,\AA.  The
\pv\ profile that we have obtained using this less than optimal
procedure is very similar to that presented by \citet{Hubetal12} in
the bottom panel of their Fig.~1. In particular, the \pv\ profiles of
all Balmer lines from H$\epsilon$ down to H10 show a negative
peak of $\sim -0.40$\,\%, and the \pv\ profile of the \ion{Ca}{ii}\,K
line shows a negative peak of $\sim -0.7$\,\%. Our H$\delta$ line
shows a milder polarimetric feature than that found by
\citet{Hubetal12}, with a negative peak of $\sim - 0.3$\,\% instead of
$\sim -0.4$\,\%.  Our corresponding \bz\ value obtained from H Balmer
lines is $-325 \pm 105$\,G, to be compared with their $-400 \pm
50$\,G.

Note that the bottom-left panel of our Fig.~\ref{Fig_HD92207_spikesVN}
shows that the linear fit exhibits a very uneven/skewed point
distribution. This fully reflects the fact that the observed Stokes
$V$ signal is not due to Zeeman effect (which would be responsible for
a \pv\ signal symmetric about the interpolating line). We note that
our smallest error bar is larger than that found by \citet{Hubetal12};
this discrepancy is probably due to the fact that we have rescaled our
error bar with the square-root of the reduced $\chi^2$, as explained
in \citet{Bagetal12}; indeed, by simply using the internal error
  bar, i.e., without re-scaling with the $\chi^2$, our error bar would
  be the same as that published by \citet{Hubetal12}, and in both cases of
Fig.~\ref{Fig_HD92207_spikesVN} and \ref{Fig_HD92207_nospikesN} we would 
obtain a $6\,\sigma$ detection.

The null profile of our Fig.~\ref{Fig_HD92207_spikesVN} is totally
different from that published by \citet{Hubetal12}: our \nv\ profile
shows several spikes not visible\ in \citet{Hubetal12}, and, similarly
to the case of Fig~\ref{Fig_HD92207_nospikes}, leads to a $4.7\,\sigma$ field
detection ($\nz=-355\pm75$\,G). The reasons of this puzzle are
investigated in the next Section.

\subsection{The non-uniqueness of the null profiles}\label{Sect_Null}
\begin{figure}
\begin{center}
\scalebox{0.58}{
\includegraphics*[2.5cm,9.1cm][20cm,20.4cm]{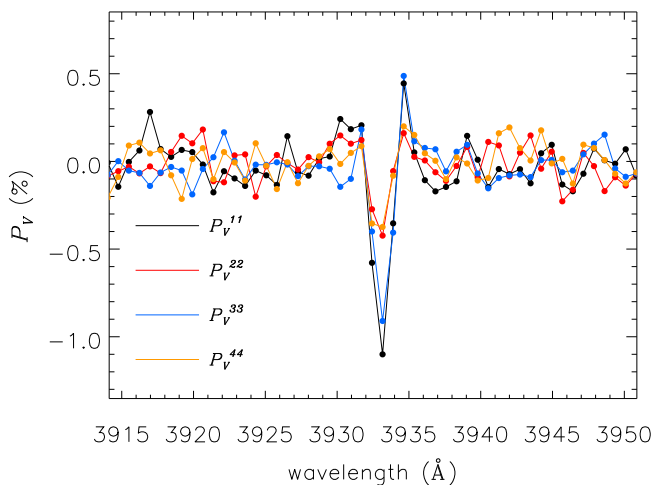}}
\end{center}
\caption{\label{single_stokesV} Single rectified $\pv^{jj}$\ profiles, in the
region of the \ion{Ca}{ii}\,K line, obtained using a slightly imperfect
wavelength calibration. The profiles are numbered as in
Fig.~\ref{Fig_CaII_Cal_2011}. Note that this plot was obtained with a
less accurate wavelength calibration than the plot shown in the rightmost
panel of Fig.~\ref{Fig_CaII_Cal_2011}.
}
\end{figure}

\begin{figure}
\scalebox{0.45}{
\includegraphics*[0.5cm,5.5cm][22cm,25cm]{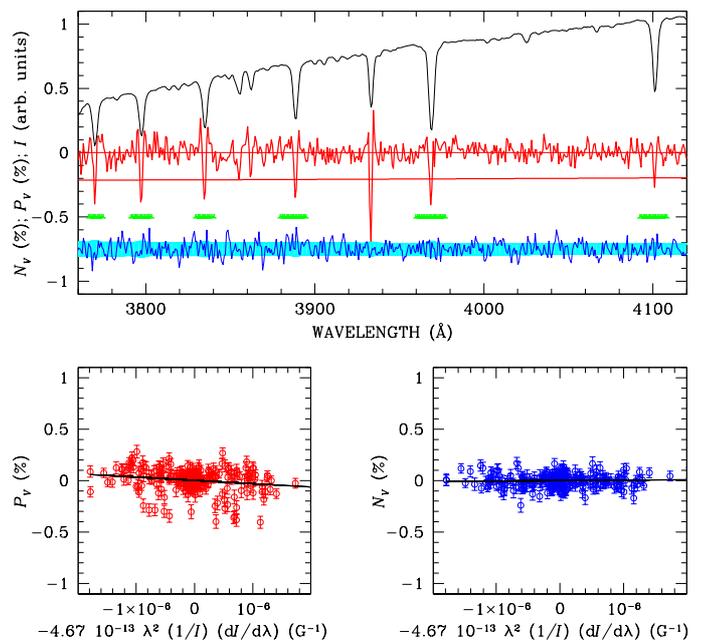}}
\caption{\label{Fig_HD92207_nospikesN}
As Fig.~\ref{Fig_HD92207_spikesVN}, but reshuffling the order of the files
as explained in the text. Spikes are now visible only in the \pv\ profile, while the
\nv\ profile appears smoother than in Fig.~\ref{Fig_HD92207_spikesVN}, and
the corresponding \nz\ values is now $45\pm65$\,G.}
\end{figure}

Null profiles have been introduced by \citet{Donetal97}, and their
definition and meaning were discussed by \citet{Bagetal09}. In the
context of the double difference method, the null profiles can be
regarded simply as the difference between the \pv\ profiles
obtained from the combination of different pairs of frames.
Let us assume that $i$ labels the frames obtained
with the retarder waveplate at $-45\degr$, and $j$ labels the frames
obtained with the retarder waveplate at $+45\degr$, and that
these indices run in a temporal sequence (i.e., $i=2$ is taken
after $i=1$ and so on). In case of HD\,92207, we have $i=1, \ldots, 4$ and $j=1,
\ldots, 4$, hence we can calculate sixteen different $\pv^{ij}$
profiles (although only four totally independent of each other); more
in general, from $N$ pairs of frames one can obtain $N^2$ individual
$\pv^{ij}$ profiles. While their average \pv\ profile is unique,
the definition of the null profile, if it is not explicitly linked to a
temporal sequence, remains ambiguous,
as one may calculate\\ 
$(1/8)\left[(\pv^{11}-\pv^{22}) + (\pv^{33}-\pv^{44}) \right]$, or \\
$(1/8)\left[(\pv^{11}-\pv^{33}) + (\pv^{22}-\pv^{44}) \right]$, or \\
$(1/8)\left[(\pv^{12}-\pv^{23}) + (\pv^{34}-\pv^{41}) \right]$,
etc. From four pairs of frames, formally we can construct 36 different \nv\
profiles, and from $N$ pairs of frames, the number of different null profiles $\nv^{ij}$
that can be calculated by reshuffling the order of the frames is
\begin{equation}
\left(\frac{N!}{\big[(N/2)!\big]^2}\right)^2 
\end{equation}
(or half of this number if consider as identical the null profiles 
that have the same absolute value). For simplicity, one
can consider only the cases where $i=j$. This way
we can still construct $N$ $\pv^{jj}$ profiles, and $N!/[(N/2)!]^2$ $\nv^{jj}$
profiles (or half this number if we consider as identical those null profiles
with identical absolute value). Figure~\ref{single_stokesV} shows the
four $\pv^{jj}$ profiles that can be obtained from the observations
of HD\,92207. Note that there are strong
similarities between the $\pv^{11}$ and $\pv^{33}$ profiles, as well as
between the $\pv^{22}$ and $\pv^{44}$ profiles.  Therefore the null profile
calculated following the temporal distribution of the frames,
i.e. $(1/8)\left[ (\pv^{11}-\pv^{22}) + (\pv^{33}-\pv^{44}) \right]$, shows a
spike. On the other hand, null profile calculated for example as
$(1/8)\left[ (\pv^{11}-\pv^{33}) + (\pv^{22}-\pv^{44}) \right]$, or \\
$(1/8)\left[ (\pv^{11}-\pv^{33}) + (\pv^{44}-\pv^{22}) \right]$ would
be much smoother, because $\pv^{11}\approx\pv^{33}$ and $\pv^{22}\approx\pv^{44}$.
To produce Fig.~\ref{Fig_HD92207_spikesVN} we have used what we consider
the most natural approach, i.e., we have kept the temporal order of the files,
and calculated
\begin{equation}
\begin{array}{rcl}
\nv &=& \frac{1}{8} \sum_{j=1}^{4} \ (-1)^j
\left[ 
\left(\frac{\fo - \fe}{\fo + \fe}\right)_{-45^\circ_j} - 
\left(\frac{\fo - \fe}{\fo + \fe}\right)_{+45^\circ_j}\right]\, \\
    &=& \frac{1}{8} \sum_{j=1}^{4} (-1)^j\, \pv^{jj} \\
\end{array}
\label{Eq_nv}
\end{equation}
where for $\alpha=-45\degr$, j=1, 2, 3, 4 correspond to the frames obtained at UT 
03:56:53,
04:01:08,
04:02:12, and
04:06:29, respectively,
and for $\alpha=+45\degr$, j=1, 2, 3, 4 correspond to the frames obtained at UT
03:58:28, 
03:59:32,
04:03:49, and
04:04:54, respectively.
By reshuffling the order of the frames as explained above, we could obtain a much
smoother \nv\ profile, qualitatively more similar to the one presented by
\citet{Hubetal12}. Figure~\ref{Fig_HD92207_nospikesN} shows the null profile obtained
from Eq.~(\ref{Eq_nv}) by imposing that for $\alpha=-45\degr$, j=1, 2, 3, 4 corresponds
to the frames obtained
at UT 
03:56:53,
04:02:12,
04:06:29, and
04:01:08, respectively,
and that for $\alpha=+45\degr$, j=1, 2, 3, 4 correspond to the frames obtained at UT
03:58:28, 
04:03:49,
04:04:54, and
03:59:32, respectively. 
The corresponding
null field value is $\nz=45\pm65$\,G.  In the case of our best
data-reduction of Fig.~\ref{Fig_HD92207_nospikes}, from the null
profile calculated after reshuffling the frames as explained above,
the field detection in the null profile becomes $\nz=90\pm65$\,G.

A similar situation would be encountered in the context of the
double-ratio method \citep[with the further ambiguity that two
  different definitions of the null profile can be given,
  see][]{Bagetal09}.  Therefore we conclude that in case of multiple
exposures, inspection to the scattering between the
$\pv^{jj}$\ profiles obtained from different pairs of frames (or even
better the standard error of the mean normalised by the photon-noise error
bar) may be a quality check more effective than, or at least
complementary to the analysis of the null profiles. For an alternative
and more sophisticated approach involving a $\sigma$-clipping
algorithm of the $\pv^{ij}$ profiles we refer to \citet{Bagetal06}.

\subsection{High resolution spectropolarimetry of HD~92207}
Four high-resolution spectropolarimetric observations of HD~92207 were
obtained in February 2013 with the HARPSpol instrument
\citep{Piskunov2011} at the ESO 3.6-m telescope. These circular
polarisation spectra, collected in the context of the ESO programme
090.D-0256(A), cover the 3800--6910~\AA\ wavelength domain at the
resolution of $\lambda/\Delta\lambda = 109\,000$. The typical S/N
ratio of these data is 300--400 per pixel, measured at $\lambda\approx
5200$~\AA.

Each Stokes $V$ observation consisted of a series of 4 subexposures,
obtained with the quarter-wave retarder plate angles of 45\degr,
135\degr, 225\degr, and 315\degr\ relative to the beamsplitter. The
spectra were extracted using the dedicated HARPSpol pipeline based on
the {\sc reduce} code of \citet{Piskunov2002}. Individual beams were
normalised to the continuum and then combined to produce
the Stokes $V$ and diagnostic null spectra using the ratio method
outlined by \citet{Bagetal09}. Further details about the reduction of
HARPSpol data can be found in \citet{Makaganiuk2011}.

\begin{figure}
\centering
\includegraphics[width=8.5cm]{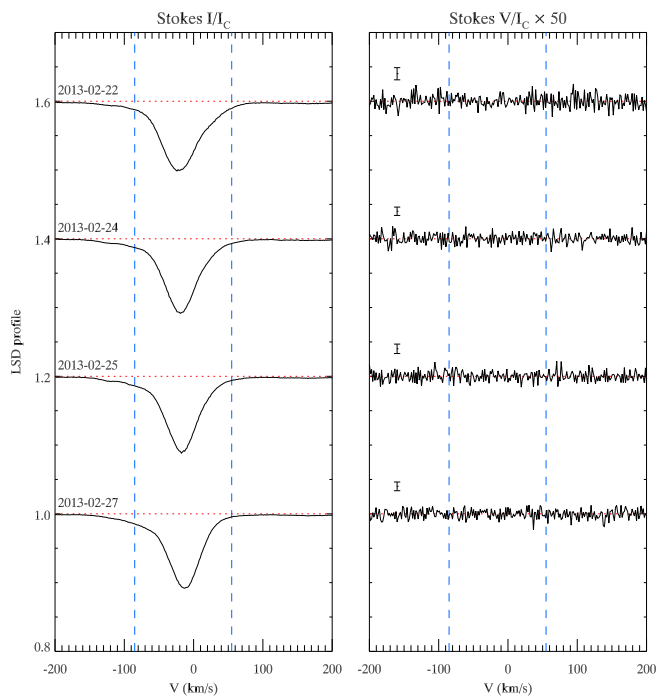}
\caption{\label{LSD} Stokes $I$ and $V$ LSD profiles derived from the HARPSpol spectra of HD~92207. Observations obtained on different nights are offset vertically. The Stokes $V$ profiles are expanded by a factor of 50 relative to Stokes $I$. The vertical dashed lines indicated the velocity range used for \bz\ calculation.
}
\end{figure}

No evidence of polarisation signatures was found in any individual
spectral lines, therefore we applied the least-squares deconvolution
method \citep{Donetal97} to obtain high precision mean Stokes $I$,
$V$, and null profiles. Using atmospheric parameters derived for
HD~92207 by \citet{Przybilla2006}, we extracted 355 metal lines deeper
than 10\,\% of the continuum from the {\sc vald} database
\citep{Kupka1999} and calculated LSD profiles with the code by
\citet{Kochukhov2010}. The resulting mean intensity and polarisation
profiles are presented in Fig.~\ref{LSD}. Despite an increase of S/N
by a factor of $\sim$\,20, no Stokes $V$ signatures are detected. The
null LSD profiles are also featureless. On the other hand, significant
night to night variation of the mean Stokes $I$ profiles can be
clearly seen.

Table~\ref{tab:harpspol} summarizes individual HARPSpol
spectropolarimetric observations of HD\,92207. The four \bz\ estimates
that we have obtained show no evidence of the magnetic field,
consistent with the lack of signatures in the LSD Stokes $V$
profiles. The typical error bar of the mean longitudinal magnetic
field derived from the high-resolution spectra is 10~G.

\begin{figure}
\centering
\includegraphics[width=8.5cm]{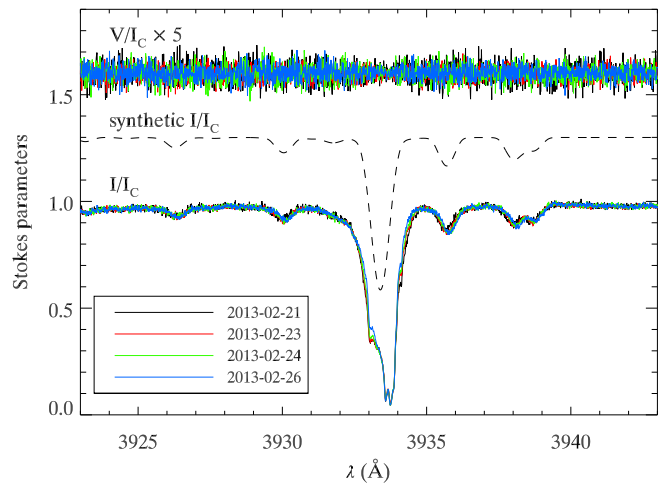}
\caption{\label{Caline} High-resolution Stokes $I$ and $V$ spectra of HD~92207 in the vicinity of the \ion{Ca}{ii} K line. Individual observations (bottom solid lines) are compared to the synthetic spectrum (mid dashed line). The latter is offset vertically by $+0.3$ for display purpose. Stokes $V/I_{\rm c}$ profiles (upper solid lines) are multiplied by a factor of 5 and offset by $+1.6$.
}
\end{figure}

Returning to the individual spectral lines, we note that the intensity
profile of the \ion{Ca}{ii} K line for which \citet{Hubetal12} reported a circular
polarisation amplitude of up to 1\,\%, exhibits a peculiar shape in our
high-resolution spectra. As illustrated by Fig.~\ref{Caline}, the
observed Ca line is noticeably broader and deeper than predicted by
the synthetic spectrum matching the stellar parameters determined by
\citet{Przybilla2006}. Furthermore, the line core is asymmetric and
shows a double structure. The \ion{Ca}{ii} H line exhibits the same
profile. Similarly, the resonance \ion{Na}{i} D lines are also very
deep and show a two-component structure. It appears that all these
strong features are formed in the circumstellar environment or in
corotating wind structures \citep{Ignetal09} rather than in the
stellar atmosphere. No circular polarisation is
detected in any of our HARPS spectra, consistently with our suggestion
that the signal measured with FORS is spurious.

\begin{figure}
\scalebox{0.62}{
\includegraphics*[3.5cm,8.8cm][22cm,20.2cm]{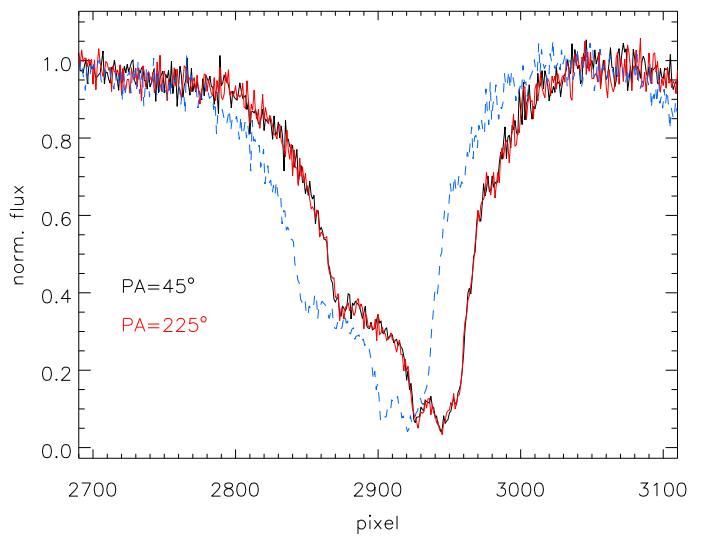}}
\caption{\label{Fig_Castable} Profiles of the \ion{Ca}{ii}\,K line extracted
from the HARPS frames of the HJD 2456347 (parallel beam). The blue line
shows how the offset observed with FORS would appear in HARPS data if it
were due to variability intrinsic to the star.
}
\end{figure}

Finally, the black and red lines of Fig.~\ref{Fig_Castable} show the
  profiles observed in the parallel beam with retarder waveplate at
  $45\degr$ and $225\degr$, respectively. These profiles appear
  constant within photon-noise error bars. The blue line (blue-shifted
  by $~24$\,px with respect to HARPS observations) shows how the offset
  observed with FORS between profiles 5 and 8 in
  Fig.~\ref{Fig_CaII_2011} would appear if observed at the spectral
  resolution of HARPS. In conclusion, HARPS data suggest that the
  offsets observed with FORS are not due to stellar variability.

\begin{table}
\caption{HARPSpol observations of HD~92207}
\centering
\begin{small}
\begin{tabular}{cccr@{\,$\pm$}lr@{\,$\pm$}lr}
\hline\hline
     HJD & S/N   & S/N  &\multicolumn{2}{c}{\bz} &\multicolumn{2}{c}{\nz} & FAP\\
-2450000 & (/px) &(LSD) &\multicolumn{2}{c}{(G)} &\multicolumn{2}{c}{(G)} & ($V$)\\
\hline
6345.6732 & 300 & 5629 & $0.2$  &$13.4$ & $  2.0$&$13.4$ & 0.897 \\
6347.6720 & 440 & 8222 & $-19.1$&$ 9.5$ & $ 18.9$&$ 9.5$ & 0.020 \\
6348.6735 & 390 & 7299 & $11.8$ &$10.7$ & $-21.4$&$10.7$ & 0.402 \\
6350.6660 & 420 & 7893 & $-1.8$ &$10.1$ & $-13.8$&$10.1$ & 0.147 \\
\hline
\end{tabular}
\end{small}
\label{tab:harpspol}
\tablefoot{Columns give the heliocentric Julian date of observations, the peak S/N ratio per pixel of Stokes $V$ spectra, the S/N achieved in the LSD profiles, and the estimates of the mean longitudinal magnetic field derived from Stokes $V$ and diagnostic null LSD profiles. The final column gives the false alarm probability that the structure in the mean circular polarisation profile can be attributed to random noise. Marginal detection corresponds to FAP\,$<10^{-3}$.}
\end{table}


\section{Discussion and conclusions}\label{Sect_Discussion}
There has been a debate in the literature regarding the reliability of
some magnetic field measurements obtained with the FORS instrument of
the ESO VLT during various surveys. In particular, there have been
numerous cases of field detection obtained with FORS that were not
confirmed by follow-up with the same or other instruments. This
situation has been discussed in detail by \citet{Bagetal12}, who have
concluded that 1) in some cases FORS data have not been fully
correctly reduced, and 2) photon-noise is not the only source of
measurement error. In particular, \citet{Bagetal12} have proposed that
very small instrument instabilities, probably unavoidable in
slit-fed Cassegrain-mounted instruments, may contribute a non-negligible
fraction of the actual field measurement error bar. In this paper we
have addressed this issue in greater detail, using simple numerical
simulations as well as calibration and science data.

Our numerical simulations show for instance that offsets as small as
1\,\% of the FWHM may lead to a substantial spurious polarisation
signal in deep spectral lines. The beam swapping technique may cancel
out such spurious effects if the offsets are both stable and
systematic (i.e., always affecting the parallel or the perpendicular
beam, or both beams obtained at a certain position angle of the
retarder waveplate, in the same way). However, certain combinations of
systematic offsets, or simply random offsets, may lead to spurious
polarisation signals even when observations are obtained with beam
swapping.

The presence of polarimetric optics (with moving parts) introduces
offsets. In FORS, the Wollaston prism is responsible for a 0.55\,px
offset between the parallel and the perpendicular beam; when the
retarder waveplate rotates from $-45\degr$\ to $+45\degr$, both beams
are blue-shifted by about 0.13\,px (in our case $\sim
0.1$\,\AA). These combined effects cannot be compensated by the beam
swapping technique, and must be corrected by a careful wavelength
calibration.  Practically, our experience shows that while it is
mandatory to calibrate independently the parallel and the
perpendicular beam, it is safer to use the same pair of wavelength
calibrations for all frames obtained at different position angles of
the retarder waveplate. The two wavelength solutions must be obtained
adopting exactly the same fitting function and using exactly the
same arc lines. We have also noted that different offsets are
introduced when the retarder waveplate is rotated at 135\degr\ and
225\,\degr, which suggests that more accurate results may be achieved
by observing with the retarder waveplate set at two position angles
only, instead of four.

The positions of the spectra are not perfectly reproducible when the
retarder waveplate is removed and re-inserted: global offsets of $\sim
0.1$\,px have been observed when comparing calibration data obtained
in different days. This suggests that, to achieve the highest
accuracy, scientific frames can be safely combined only if the
observing series has not been interrupted to remove and re-insert the
retarder waveplate (or any other optical component, such as the
grism).  If the instrument setting is changed before the series is
completed (other than by the rotation of the retarder waveplate), we
recommend that the entire series be repeated.  We are not able to
assess whether random offsets may occur following a rotation
of the retarder waveplate, but we found no evidence of this. We note
that a potential global offset between daytime calibrations and night
time scientific frames does not represent a problem because the offset
between parallel and perpendicular beam remains constant, and may be
corrected.

We have found strong evidence for small and presumably random offsets
in the FORS observations of HD\,92207, a supergiant A0 star in which
\citet{Hubetal12} have recently claimed a field detection, and that we
have adopted here as a case study. Spectra from the same beam,
obtained with the retarder waveplate in identical positions, appear
shifted by up to $\sim 0.25$\,px from each other. In the context of
ultra-high signal-to-noise ratio spectro-polarimetric data, these
random offsets are not negligible and contribute significantly to the
error budget. However, the actual magnitude of the observed offsets is
very small ($\sim 0.03\,\arcsec$), hence it is not easy to pin down
their origin. While they might be linked to the non reproducibility of
the position of the spectral lines as the retarder waveplate moves, they might
well have a completely different origin. In general, they could be due
to the star itself (rapid pulsation or rapid changes of the radial
velocity), but this (unlikely) hypothesis seems ruled out completely
by the fact that similar offsets are not present in our HARPS
data. They could be due to differential instrument flexures \citep[see
  a detailed discussion in][]{Bagetal12}, and/or to movements of the
stellar image in the slit due either to seeing or to guiding
inaccuracy. Actually, since FORS observations of HD\,92207 were
obtained under excellent seeing conditions ($\sim 0.7\arcsec$) and
with extremely short integration time ($t = 3$\,sec each exposure),
image movements due to atmospheric variability are likely. Another
potential cause is a time dependent differential atmospheric
diffraction. During the observations, the slit was nearly
perpendicular to the parallactic angle, and between the beginning and
the end of the exposures, the airmass varied from 1.55 to
1.65. Airmass 1.55 is the limit up to which the Longitudinal
Atmospheric Dispersion Corrector \citep{Avietal97} may fully compensate.
However, we could not confirm that the observed offset is
wavelength dependent, as is atmospheric dispersion. Clearly, the
observed offsets may be due to a combination of all the effects
mentioned above. More insight could be provided with the help of a
systematic investigation into the archive data, and few technical
tests (e.g., measuring the stability of the centroid of the star
imaging obtained through the slit, with no grism inserted; obtaining a
series of wavelength calibration with the telescope tilted, while the
instrument is rotating).

How do these instabilities affect the reliability of FORS measurements,
and how can their impact be evaluated?

Null profiles may or may not reveal instrumental problems, and we
should recall that they have mainly a statistical meaning. Using the
longitudinal field values \nz\ calculated from the null profiles for a
very large sample of FORS1 data, \citet{Bagetal12} concluded that,
statistically, the actual field error bars are up to 50\,\% larger than
those calculated from photon-noise only. However, when looking at an individual
measurement, an \nz\ value consistent with zero does not automatically
imply that the corresponding \bz\ value is not altered by instrumental
(or data-reduction) effects. Figure~\ref{Fig_HD92207_nospikesN} is a
convincing example of how a combination of instrument instabilities
and non-perfect wavelength calibration leads to a number of circular
polarisation features associated with a clean null profile.

In this paper we have also shown that if the observing series
comprises more than two pairs of frames, the definition of null
profile does not have a unique interpretation, and null profiles
obtained by sorting the frames in different ways are
different. Therefore, especially in case of dubious detections, we
recommend calculating and inspecting the scatter of the \pv\ profiles
obtained from different pairs of frames (e.g., one could plot the
ratio between the standard error of the mean and the photon-noise error bar). Another
important test is to compare the spectra recorded in the same beam and
obtained with the retarder waveplate at identical position, and to
check if they are consistent within photon-noise error bars.

In spite of all the issues discussed in this paper, we should remark
that if data are treated correctly, spurious detections are relatively
unlikely to occur. In our case study, we could reproduce a previously
reported field detection only by degrading the accuracy of our
wavelength calibration. Even in that case, applying a $\sigma$
clipping algorithm to the least-square fit would bring our field value
to less than 3\,$\sigma$ from zero provided the external error is
  adopted as uncertainty estimate. Finally, we recall that the spikes
of the \pv\ profile shown in Fig.~\ref{Fig_HD92207_nospikesN} should
not be mistaken for Zeeman signatures. A genuine Zeeman signature
should exhibit a typical S profile (except when longitudinal field is
reversing sign and the star rotates rapidly, but this cross-over
effect can be revealed only at higher spectral resolution), and the
zero-order moment of Stokes $V$ about line centre should always be
zero.

We should also note that in this paper we have made no attempts to
establish how frequently instabilities occur at a significant
level. Accordingly, there is no implication that the offsets at the
level detected in the observations of HD\,92207 are typical of FORS1/2
measurements.

In fact, the considerations presented in this paper are not restricted
to the FORS instrument, but are of a much more general relevance. It
is not unusual that instruments are used for tasks for which they were
not originally designed, and in these cases it is up to the user to
demonstrate that instrumental sensitivity corresponds to its
accuracy. Calculating the square-root of the photon count is a very
simple and fundamental task, but not necessarily sufficient to
evaluate the reliability of astronomical measurements.

\begin{acknowledgements}
We thank Naum Rusomarov for performing the HARPSpol observations of HD\,92207.
JDL acknowledges financial support for this project from the Natural Sciences
and Engineering Research Council of Canada.
OK is a Royal Swedish Academy of Sciences Research Fellow, supported by grants
from Knut and Alice Wallenberg Foundation and Swedish Research Council.
We thank the anonymous referee, Gregg Wade 
for very useful comments to the manuscript.
\end{acknowledgements}


\begin{thebibliography}{}
\bibitem[Alecian et al.(2013)]{Aletal13}
        Alecian, E., Wade, G.A., Catala, C., et al. 2013, \mnras, 429, 1001
\bibitem[Appenzeller et al.(1998)]{Appetal98} 
	Appenzeller, I., Fricke, K., Furtig, W., et al. 1998, The Messenger, 94, 1
\bibitem[Auri\`ere et al.(2009)]{Auretal09} 
        Auri\`ere, M., Wade, G.A., Konstantinova-Antova, R. et al. 2009, \aap, 504, 231
\bibitem[Avila et al.(1997)]{Avietal97} Avila, G., Rupprecht, G., \& Beckers, J.M.
        1997, SPIE, 2871, 1135
\bibitem[Bagnulo et al.(2002)]{Bagetal02}
	Bagnulo, S., Szeifert, T., Wade, G.A., Landstreet, J.D., \& Mathys, G. 2002, \aap, 389, 191
\bibitem[Bagnulo et al.(2006)]{Bagetal06} 
	Bagnulo, S., Landstreet, J. D., Mason, E., et al. 2006, \aap, 450, 777
\bibitem[Bagnulo et al.(2009)]{Bagetal09} 
	Bagnulo, S., Landolfi, M., Landstreet, J.D., Landi Degl'Innocenti, E., Fossati, L., \& Sterzik, M. 2009, \pasp, 121, 993
\bibitem[Bagnulo et al.(2012)]{Bagetal12} 
	Bagnulo, S., Landstreet, J.D., Fossati, L., \& Kochukhov, O. 2012, \aap, 538A, 129
\bibitem[Catala et al.(2007)]{Catetal07}
        Catala, C., Alecian, E., Donati, J.-F., et al. 2007, A\&A, 462, 293 
\bibitem[Donati et al.(1997)]{Donetal97}
	Donati, J.-F, Semel, M., Carter, B.D., Rees, D.E., \& Collier Cameron, A. 1997, \mnras, 291, 658
\bibitem[Donati \& Landstreet(2009)]{Donetal09}
        Donati, J.-F., \& Landstreet, J.D. 2009, ARA\&A, 47, 333 
\bibitem[Grunhut et al.(2010)]{Gruetal10}
        Grunhut, J.H., Wade, G.A., Hanes, D.A., \& Alecian, E. 2010, 408, 2290
\bibitem[Grunhut \& Wade(2012)]{GruWad12}
        Grunhut, J.H., \& Wade, G.A. 2012, ASPC, 465, 42
\bibitem[Henrichs et al.(2000)]{Henetal00} Henrichs, H., de Jong,
  J.\ A., Donati, J.-F., et al. 2000, in ``The Be Phenomenon
  in Early-Type Stars'', ed. M.\ A. Smith, H.\ F. Henrichs, J.\ Fabregat,
  ASP Conference Series, 214, 324
\bibitem[Hubrig et al.(2012)]{Hubetal12} 
	Hubrig, S., Schoeller, M., Kholtygin, A.F., Gon\'zalez, J.F., Kharchenko, N.V., 
        \& Steffen. M. 2012, \aap, 546, L6 
\bibitem[Ignace et al.(2009)]{Ignetal09} 
	Ignace, R., Hubrig, S., \&  Sch\"{o}ller, M. 2009, \aj, 137, 3339
\bibitem[Izzo et al.(2010)]{Izzetal10} Izzo, C., de Bilbao, L., Larsen, J., et al. 2010,
        SPIE, 7737, 773729
\bibitem[Jordan et al.(2012)]{Joretal12}
        Jordan, S., Bagnulo, S., Werner, K., \& O’Toole, S.J.O. 2012, A\&A 542, A64
\bibitem[Kochukhov et al.(2010)]{Kochukhov2010} 
  	Kochukhov, O., Makaganiuk, V., \& Piskunov, N. 2010, \aap, 524, A5
\bibitem[Kupka et al.(1999)]{Kupka1999} 
	Kupka, F., Piskunov, N., Ryabchikova, T.~A., Stempels, H.~C., \& Weiss, W.~W. 1999, \aaps, 138, 119
\bibitem[Landstreet et al.(2012)]{Lanetal12}
        Landstreet, J.D., Bagnulo, S., Fossati, L., Jordan, S., \& O’Toole, S.J.O. 2012,
        A\&A 541, A100
\bibitem[Makaganiuk et al.(2011)]{Makaganiuk2011} 
  	Makaganiuk, V., Kochukhov, O., Piskunov, N. et al. 2011, \aap, 525, A97
\bibitem[Neiner et al.(2003)]{Neietal03} Neiner, C., Hubert, A.-M.,
  Fr\'{e}mat, Y., et al. 2003, A\&A, 409, 275
\bibitem[Piskunov \& Valenti(2002)]{Piskunov2002}
  	Piskunov, N.~E., \& Valenti, J.~A. 2002, \aap, 385, 1095
\bibitem[Piskunov et al.(2011)]{Piskunov2011} 
  	Piskunov, N., Snik, F., Dolgopolov, A. et al. 2011, The Messenger, 143, 7
\bibitem[Przybilla et al.(2006)]{Przybilla2006}
  	Przybilla, N., Butler, K., Becker, S. R., \& Kudritzki, R.P. 2006, \aap, 445, 1099
\bibitem[Silvester et al.(2009)]{Siletal09}
        Silvester, J., Neiner, C., Henrichs, H.\ F., et al. 2009 MNRAS, 398, 1505
\bibitem[Sterzik et al.(2012)]{Steretal12}
        Sterzik, M.F., Bagnulo, S., \& Pall\'e, E. 2012, Nature, 483, 64
\bibitem[Wade et al.(2005)]{Wadetal05}
        Wade, G.\ A., Drouin, D., Bagnulo, S., et al. 2005, A\&A, 442, 31L
\bibitem[Wade et al.(2007)]{Wadetal07}
        Wade, G.\ A., Bagnulo, S., Drouin, D., Landstreet, J.\ D., \& Monin, D.
        2007, MNRAS, 376, 1145
\end{thebibliography}
\end{document}